\documentclass[preprint,showpacs,preprintnumbers,amsmath,amssymb]{revtex4}

\usepackage{graphicx}% Include figure files
\usepackage{dcolumn}% Align table columns on decimal point
\usepackage{bm}% bold math

\begin{document}

\title{The shower size parameter as estimator of extensive air
shower energy in fluorescence telescopes}

\author{Vitor de Souza}
 \email{vitor@astro.iag.usp.br}
 \affiliation{Instituto de Astronomia, Geof\'{\i}sica e Ci\^encias Atmosf\'ericas, Universidade de S\~ao Paulo}
\author{Federico Sanchez}
 \email{federico.sanchez@mi.infn.it}
 \affiliation{Dipartimento di Fisica, Universit\'a degli Studi di Milano e I.N.F.N.}
\author{Gustavo Medina-Tanco}
 \email{gmtanco@gmail.com}
 \affiliation{Instituto de Astronomia, Geof\'{\i}sica e Ci\^encias Atmosf\'ericas, Universidade de S\~ao Paulo}
\author{Jeferson A. Ortiz}
 \email{jortiz@astro.iag.usp.br}
  \affiliation{Instituto de Astronomia, Geof\'{\i}sica e Ci\^encias Atmosf\'ericas, Universidade de S\~ao Paulo}

\date{November 14th, 2005 - J.A.Ortiz }

\begin{abstract}

The fluorescence technique has been successfully used to detect
ultrahigh energy cosmic rays by indirect measurements. The
underlying idea is that the number of charged particles  in the
atmospheric shower, i.e, its longitudinal profile, can be
extracted from the amount  of emitted nitrogen fluorescence light.
However the influence of shower fluctuations  and the very
possible presence of different nuclear species in the primary
cosmic ray  spectrum makes the estimate of the shower energy from
the fluorescence data analysis a difficult task. We investigate
the potential of shower size at maximum depth as estimator of
shower  energy. The detection of the fluorescence light is
simulated in detail and the reconstruction biases  are carefully
analyzed. We extend our calculations to both HiRes and EUSO
experiments.  This approach has shown some advantages to the
reconstruction of the energy when compared to the standard
analysis procedure.

\end{abstract}


\maketitle

\section{Introduction}

Cosmic rays (CR) are the highest energy particles present in nature with
energies exceeding 10$^{20}$eV . Their origin, nature
and possible acceleration mechanisms are still a mystery, despite the
efforts of many cosmic ray experiments in the last decades. Such challenge
is in part due to the very low flux of high energy and the consequent few
observed events to be analyzed.

Measuring extensive air showers (EAS) is currently the only way to
study the cosmic ray spectrum and chemical composition at energies
around and above PeV. At energies $E$$\geq$10$^{17}$~eV the shower
development can be directly observed by measuring the fluorescence
light from atmospheric nitrogen that is excited by the ionization
of the secondary charged shower particles (essentially electrons
and positrons). Experiments applying this technique can determine
the depth of maximum air shower development ($X_{\mathrm{max}}$)
and the corresponding number of charged particles
($N_{\mathrm{max}}$). Presently, the HiRes~\cite{bib:hires}, Pierre Auger
Observatory~\cite{bib:auger-nim}, EUSO~\cite{bib:euso-artigo},
OWL~\cite{bib:owl-artigo} and  Telescope
Array~\cite{bib:telescope:array})
experiments are using or planning to use fluorescence detectors to
investigate the ultra high energy cosmic rays.

The total amount of emitted fluorescence light in a shower is a very good
approximation to the total number of charged particles $N$($X$), where $X$
is the atmospheric depth. In this sense the number of particles at shower
maximum can serve as an estimator of the shower energy. The total energy that
goes into electromagnetic charged particles is obtained by integration of
the shower longitudinal profile
\begin{equation}
E_{\rm em}=\alpha \int_0^{\infty} N(X)dX
\label{eq:Eem}
\end{equation}
where $\alpha$ is the average ionization loss rate \cite{bib:song},
and the integral on the right-hand side represents the total track
length of all charged particles in the shower projected onto the
shower axis.

As an alternative proposal \cite{bib:bruce:gap} the
electromagnetic energy can also be  calculated by using the
fluorescence light intensity and the fluorescence efficiency,
without the need of reconstructing the number of particles as a
function of the  atmospheric depth. Such approach is taken as a
very precise measurement of the primary  shower energy because it
is supposed to be weakly dependent on the simulation models  and
on the primary particle type. However, when the shower development
details are  taken into account the calorimetric measurement can
lead to high systematic uncertainties. Of no less concern is the
important fact that the fluorescence efficiency as a function of
air pressure, density and humidity is only known to a certain
extent. On approach given by Eq.~\ref{eq:Eem}, the average
ionization loss rate is used in the air shower reconstruction and
hence the energy spectrum of the electron shower particles must be
known via Monte Carlo simulation.

Although the electrons and positrons constitute the majority among
the charged particles in a shower and contribute most to the
production of fluorescence light, an also important fraction of
the shower energy is carried by particles which cannot be measured
by the fluorescence technique, i.e., particles that are invisible
to fluorescence telescopes. This so called ``missing energy'' has
been calculated by Monte Carlo air shower simulation and
contributes to the uncertainties involved in the method, being
sensitive to primary composition.

Theoretical works have shown the existence of a clear relation
between the primary energy and the maximum number of particles in
the shower. Recently, Alvarez-Mu\~niz et al.~\cite{bib:alvarez}
have studied the $N_{\mathrm {max}}$ shower quantity as an
estimator for the primary shower energy, confirming the efficiency
of this technique. Such approach was analyzed for different
primary particles and energies using a fast one-dimensional
simulation program. However, telescopes particularities and
reconstruction procedures must be considered due to the
introduction of biases and fluctuations in the calculation of
$N_{\mathrm {max}}$.

The scope of this work is to explore the possibility of estimating
the primary shower energy based on $N_{\mathrm{max}}$, taking as
case studies the HiRes~\cite{bib:hires} and
EUSO~\cite{bib:euso-artigo} fluorescence experiments, i.e., ground
and space based experiments, respectively. The telescopes
particularities and the reconstruction procedures are included in
our analysis configuring a very realistic context for the
application of the technique.

\section{$N_{\mathrm {max}}$ fluctuations}
\label{sec:fluc}

Fluctuations are intrinsic to any extensive air shower and are a
cause of uncertainty in the energy reconstruction. In addition to
that, the energy of the shower is calculated without the knowledge
of the mass of the primary particle which initiated the shower.
The nature of the primary particle affects the
longitudinal development of showers and can exert influence on the
reconstruction of the air shower energy, especially when primary
photons are taken into account.

On the other hand, reconstruction procedures must relay on
simulation programs in order to relate the measurable parameters
to energy. In addition to the choice of the particular simulation
program to be used, there is a general agreement that most shower
reconstruction uncertainties originate on uncertainties on the
hadronic interaction models.

The usual procedure used by the HiRes and Auger collaborations to
reconstruct the shower energy with fluorescence telescopes
correlates the integral of the energy deposited in the atmosphere
to the total shower energy. As mentioned before, a certain amount
of the shower energy is carried away by particles which are
invisible to the fluorescence technique, i.e., muons, neutrinos
and high energy hadrons which are not converted to fluorescence
photons. Such ``missing energy'' has been estimated by Monte Carlo
simulations and shown to be dependent on the hadronic interaction
model, primary composition, shower energy and arrival direction.

In reference \cite{bib:song}, Song et al. have estimated that the fluctuations
due to the type of primary particle are about $\sim 5\%$ for
showers initiated by proton and iron nuclei and around $\sim 20\%$
if primary photons are taken into account. Meanwhile Barbosa et
al.~\cite{bib:barbosa} and Alvarez-Muniz et al.~\cite{bib:alvarez}
have calculated similar values for hadronic primaries ($\sim$5\%)
which decrease with energy. According to \cite{bib:alvarez}, the
hadronic interaction model is the main source of systematic
fluctuation in the estimation of the missing energy at the highest
energies. Due to the differences in the multiplicities of
secondary particles simulated by QGSJET01~\cite{bib:qgsjet} and 
SIBYLL2.1~\cite{bib:sibyll} hadronic interaction models, the unseen
energy calculated with QGSJET is about 50\% higher than the value
predicted by SIBYLL at $10^{20}$ eV, which can be translated into
an uncertainty of $\sim 5\%$ in the shower energy reconstruction.

The dependence of the missing energy on the energy of the primary
is a consensus among the various studies made so far. In addition,
Barbosa and co-workers ~\cite{bib:barbosa} have studied the dependence
of the missing energy on the arrival direction of the shower and
limited it to be at most 0.7\%.

Furthermore, it has been suggested in reference ~\cite{bib:alvarez} that the
discrepancies among such studies are on the order of 1-3\% and
could be explained by different hadronic interaction models for
lower energy particles. It might be relevant to mention that both
groups used different simulation programs: Song et al. and Barbosa
et al. used different versions of the well tested CORSIKA~\cite{Heck98a}, 
while Alvarez-Muniz et al. used a hybrid fast one-dimensional simulation program~\cite{Alvarez02}.

Considering all the uncertainties in the determination of the
missing shower energy listed above, one can estimate that the
total systematic fluctuation related to the primary energy
reconstruction is about 5\% to 10\%.

In order to investigate the fluctuations in the maximum number of
particles due to the natural fluctuations of the shower, the
different primary particles and the artificial fluctuations
introduced by different hadronic interaction models, we have used
the recently released CONEX~\cite{bib:conex} shower simulation
program. We have simulated 5,000 showers initiated by proton, iron
and gamma primaries at each considered energy and for the
QGSJET01 and SIBYLL2.1 interaction hadronic models. Shower have 
been simulated with the minimum cuts available, 1 GeV for hadrons 
and 1 MeV for electromagnetic particles.

\begin{figure}[t]
\begin{center}
\includegraphics[width=10cm]{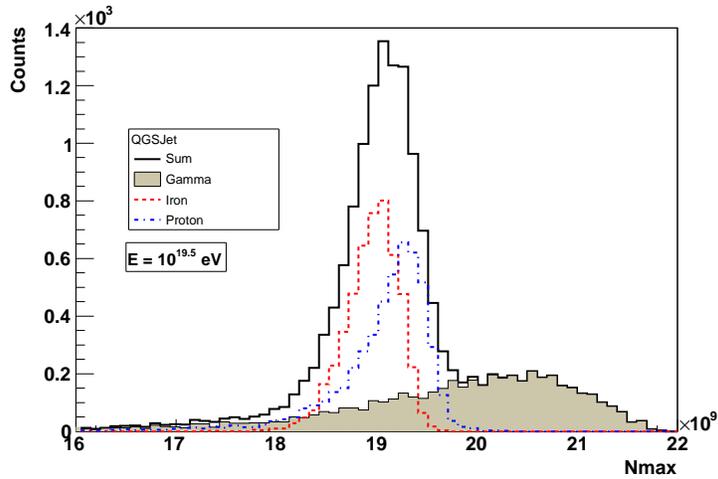}
\end{center}
\caption{Distribution of  $N_{\mathrm{max}}$ for 5,000 showers initiated by
proton, iron nuclei and gammas at $10^{19.5}$ eV. Simulation has been
done with the CONEX program. The thick solid line shows
the sum of the $N_{\mathrm{max}}$ distributions for proton, iron
nuclei and gamma primaries.} 
\label{fig:nmax:particle}
\end{figure}

Fig.~\ref{fig:nmax:particle} shows the $N_{\mathrm{max}}$
distribution for 5,000 showers initiated by protons, iron nuclei
and gamma at $10^{19.5}$ eV simulated with the QGSJET hadronic model. The
distribution illustrates the previous discussion regarding the
fluctuations due to the primary particle type. If only proton and
iron nuclei are considered, the $N_{\mathrm{max}}$ distribution
shows a very narrow distribution with median value of $1.90 \times
10^{10}$ and a dispersion of 3.6\% at the 68\% of confidence level.
If gamma shower are considered, the median of the combined
distribution (gamma+proton+iron nuclei) 
increases slightly to $1.92 \times 10^{10}$ while the dispersion
reaches 7\% at 68\% of confidence level.

The same calculation has been done for proton and iron nuclei at
$10^{20.5}$ eV and the fluctuations presented the same level of
3.6\%. Gamma air showers have not been studied at the energy of
$10^{20.5}$~eV due to the fact that the CONEX program does not
include the pre-shower algorithm
effect~\cite{bib:pre:shower,bib:pre:shower:2} that takes into
account photon interactions with the geomagnetic field affecting
the longitudinal development of showers with energy above
$10^{19.5}$~eV.

\begin{figure}[t]
\begin{center}
\includegraphics[width=10cm]{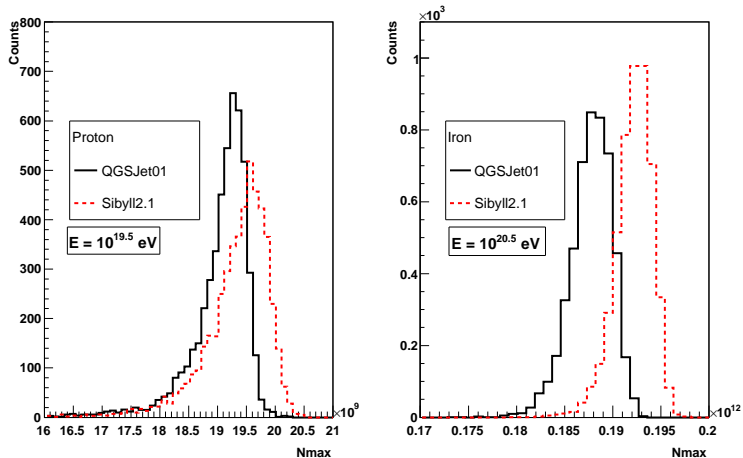}
\end{center}
\caption{Distribution of  $N_{\mathrm{max}}$ for 5,000 showers
  simulated by QGSJET and SIBYLL.  Simulation has been
done with the CONEX program.
The left hand side shows the primary protons at $10^{19.5}$ eV while the right hand side
illustrates the primary iron nuclei at the energy of $10^{20.5}$~eV.}
\label{fig:nmax:modelo}
\end{figure}

In order to investigate the uncertainty due to the hadronic
interaction model we have simulated 5,000 proton showers with the CONEX
program using SIBYLL and
QGSJET hadronic interaction models.  Fig.~\ref{fig:nmax:modelo}
illustrates the 
$N_{\mathrm{max}}$ distribution for proton showers simulated with
QGSJET and SIBYLL at $10^{19.5}$ eV and the corresponding 
fluctuation is  4.6\%. Fig.~\ref{fig:nmax:modelo} also shows
the same distribution for iron nuclei at  $10^{20.5}$ eV and the
corresponding fluctuation is 3.3\%.

Comparing the numbers given above and Figs.~\ref{fig:nmax:particle}
and \ref{fig:nmax:modelo}, we are able to conclude that the
systematic uncertainties in $N_{\mathrm{max}}$ are dominated by
the hadronic interaction models rather than mass composition if proton
and iron nuclei are considered, however the fluctuations due to both
effects are very similar, 4.6\% and 3.6\%, respectively at $10^{19.5}$
eV. If 
primary gamma are considered, 
mass composition would be more relevant than hadronic interaction
models to the $N_{\mathrm{max}}$ fluctuations. 

\begin{figure}[t]
\begin{center}
\includegraphics[width=10cm]{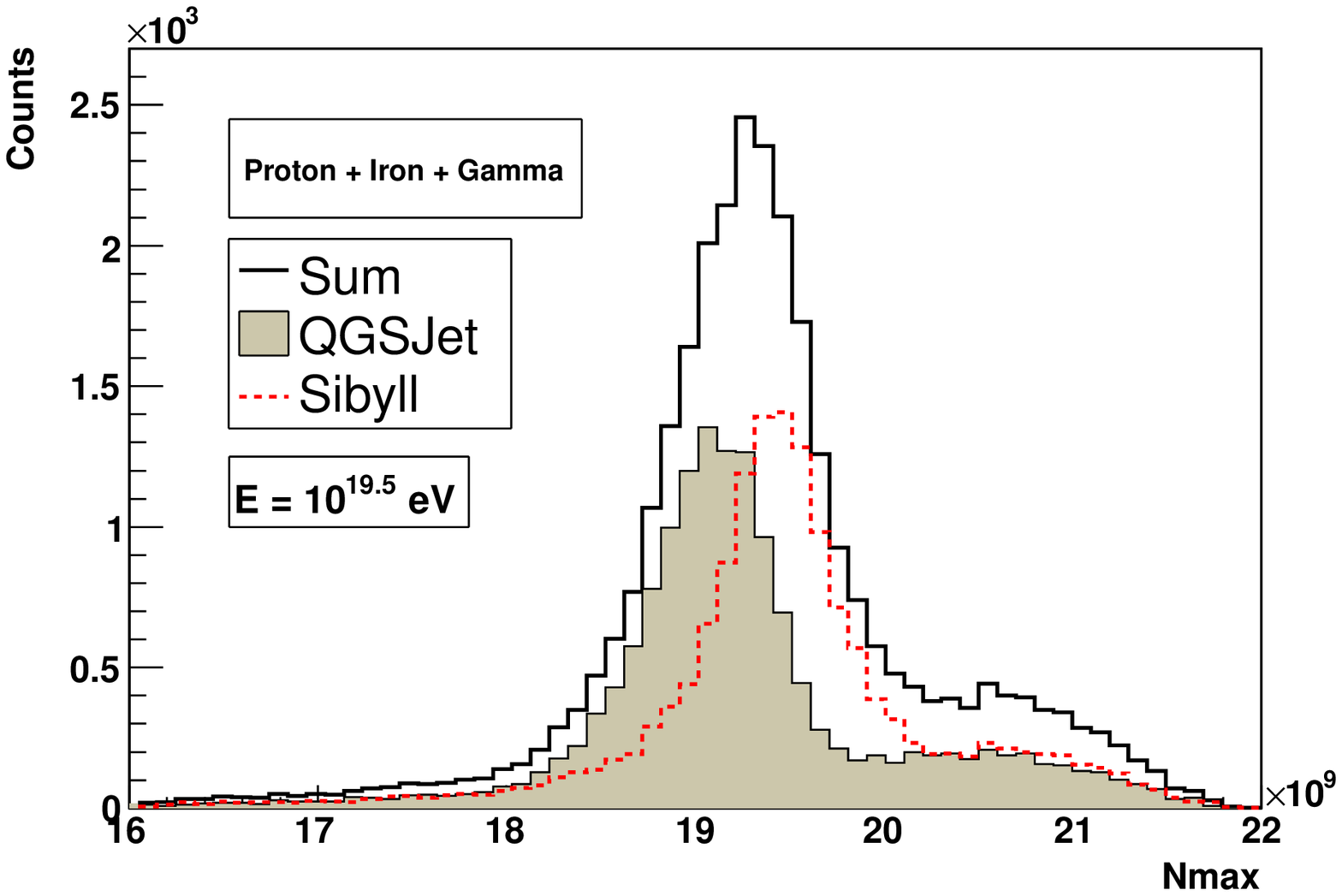}
\end{center}
\caption{Distribution of  $N_{\mathrm{max}}$ for 5,000 showers
  simulated by QGSJET and SIBYLL and initiated by 
proton, gamma and iron nuclei at $10^{19.5}$ eV . Simulation has been
done with the CONEX program.}
\label{fig:nmax:tudo}
\end{figure}

In order to exemplify the increase in the uncertainty by allowing
the introduction of photons in the cosmic ray flux, we also
consider the peculiar case of equal abundances of protons, iron
and gammas. Fig.~\ref{fig:nmax:tudo} shows 5,000 air showers
induced by gamma, proton and iron nuclei simulated with QGSJET and
SIBYLL hadronic interaction models, at the energy of
$10^{19.5}$~eV. The total $N_{\mathrm{max}}$ uncertainty is $\sim 7\%$
which is 
comparable to the systematic uncertainties introduced by the
missing energy correction calculation.

%##############################################################################
%##############################################################################
%##############################################################################
%##############################################################################
%##############################################################################
%##############################################################################

\section{Shower and Telescope Simulation}
\label{sec:sim}

The results presented in the current section have been obtained by
using the CORSIKA 6.2 simulation program. In addition, gamma
showers have been simulated by considering the pre-shower effect.
The thinning factor \cite{Hillas97a} of the simulations is $10^{-6}$ and the
longitudinal air shower profiles were sampled in steps of 5
g/cm$^2$. The energy thresholds were set to 0.1 MeV, for electrons
and photons, 0.3 GeV for hadrons and 0.7 GeV for muons. Protons
and gamma ray showers were simulated in the energy interval from
$10^{19}$ to $10^{20.5}$ eV in steps of $0.1$~dex. For each energy
and primary particle we simulated 100 events which were recycled
50 times by randomly drawing a new arrival direction and core
position. For gamma showers simulated with the pre-shower effect
only the core position was randomized since the relative direction
of the primary particle with the geomagnetic field affects the
development of the shower.

The simulations of the HiRes telescopes were performed following
the same general procedure adopted by the HiRes Collaboration as
published in \cite{bib:hires:espectro} and
\cite{bib:hires:espectro:2}. Comparisons of our simulation with
the HiRes Collaboration simulation and the HiRes data can be seen
in \cite{bib:espectro:2005}.

A similar simulation program was written for the EUSO telescope in
which we incorporated the geometry details of that telescope. The
EUSO experiment has been proposed to operate at 430~km of altitude
looking downwards to measure the fluorescence light produced by
the passage of a cosmic ray shower. It is configured with a 60
degree field of view covered by pixels of 0.1 degree resulting in
200.000~km$^2$ detection area.

For the HiRes studies we have simulated shower cores within a
radius of 50~km from the telescope. The zenith angle has been
randomly chosen from 0 to 60 degrees. For the EUSO studies we
allowed showers with cores within a radius of 250~km from the axis
of the telescope and the zenith angle was chosen from 0 to 90
degrees.

\begin{figure}[t]
\begin{center}
\includegraphics[angle=-90,width=10cm]{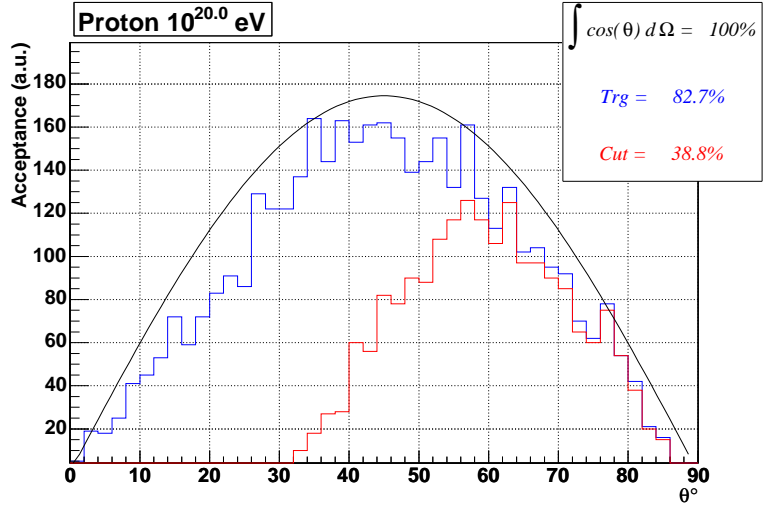}
\end{center}
\caption{EUSO acceptance for proton showers at $10^{20}$~eV as a
function of the zenith angle according to our simulation program.
The heavy smooth line shows the isotropic distribution of events,
the full line histogram corresponds to the events which trigger
the telescope and the light line histogram the events which
survived to the quality cuts.}
%VI-COM
% discutir com Gustavo e Jeferson se devemos colocar este grafico.
% pedir figura pro Federico com linhas tracejadas ao inves de coloridas.
\label{fig:euso:accep}
\end{figure}

Besides the configuration details, included in our simulation
program, we also paid careful attention to the simulation of the
atmosphere since the field of view of the EUSO telescope spans the
entire atmosphere. In order to verify our code we have the EUSO
acceptance as a function of the arrival direction of the primary
particle. Fig.~\ref{fig:euso:accep} illustrates the acceptance for
the EUSO telescope, which is in good agreement with the predicted
values given in \cite{bib:euso-artigo}.

\begin{figure}[t]
\begin{center}
\includegraphics[width=10cm]{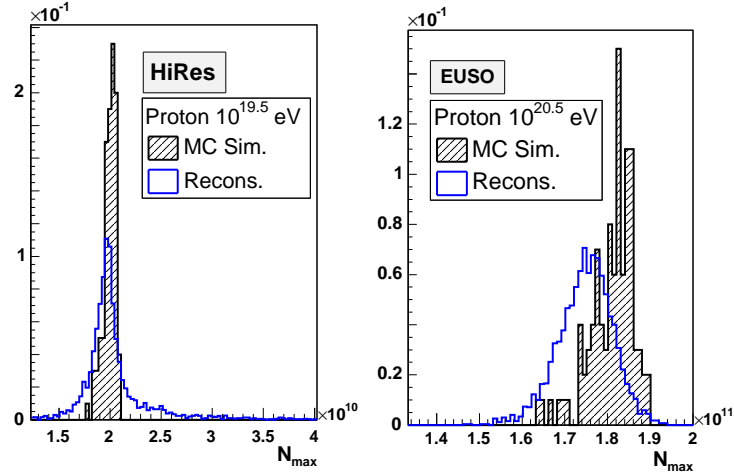}
\end{center}
\caption{Comparison between the reconstructed and simulated $N_{\mathrm{max}}$
for the HiRes-I and EUSO telescopes.}
\label{fig:nmax:rec:sim}
\end{figure}

The importance of a detailed telescope simulation in the studies
of $N_{\mathrm{max}}$ as an energy estimator can be seen in
Fig.~\ref{fig:nmax:rec:sim} in which the simulated
$N_{\mathrm{max}}$ distribution as predicted by the CORSIKA code
is compared to the reconstructed $N_{\mathrm{max}}$, obtained
after the detection by the EUSO and HiRes-I telescopes. It is
evident from the figure that the distortion of the
$N_{\mathrm{max}}$ distribution after the detection and
reconstruction of the shower must be taken into account properly.
Fluorescence telescopes have detection biases regarding arrival
direction, primary energy and core position which are very
difficult to control. Moreover, the reconstruction procedure also
introduces fluctuations which depend on several parameters. These
effects result in a distortion of the $N_{\mathrm{max}}$
distribution that depends on the configuration of the telescope
and the chosen reconstruction method. Therefore the simulation of
the telescope and the application of the reconstruction procedure
is fundamental in order to obtain a meaningful relation between
$N_{\mathrm{max}}$ and energy and to properly estimate the
reconstruction errors attained by this method.

%##############################################################################
%##############################################################################
%##############################################################################
%##############################################################################

\section{Primary Energy Reconstruction}

For the energy reconstruction of a shower there is one procedure
which is unavoidable by all proposed methods, especially those in
investigation here. Each pixel hit in the telescopes corresponds
to a given atmospheric depth and the respective signal measured by
it can be correlated to the number of particles or, more directly,
to the energy deposited by the shower at that depth. From a
limited number of points along the longitudinal profile, a
function must be fitted in order to calculate, by extrapolation,
the energy deposited by the shower beyond the limits of the field
of view of the telescope. The most widely used function is the
Gaisser-Hillas, which efficiency have been extensively tested and
confirmed~\cite{bib:hires:espectro}.

In this section, we are going to investigate the influence of the
fitting procedure in the uncertainty of the energy for two
reconstruction methods. The analysis will be done for a ground
based experiment and may differ for a space based telescope.

The standard method used by the HiRes and Auger Collaborations is
based on the integration of the fitted Gaisser-Hillas function to
obtain the shower energy and, from now on, it will be referred as
the integral method. In addition, the method based on the
$N_{\mathrm{max}}$ parameter will be referred as the
$N_{\mathrm{max}}$ method. Our intention here is to evaluate the
influence of the knowledge of a limited portion of the shower
longitudinal development in the determination of the parameters
related to energy, which are the integral of the longitudinal
profile for the integral method and the $N_{\mathrm{max}}$
parameter for the $N_{\mathrm{max}}$ method.

\begin{figure}[t]
\begin{center}
\includegraphics[width=10cm]{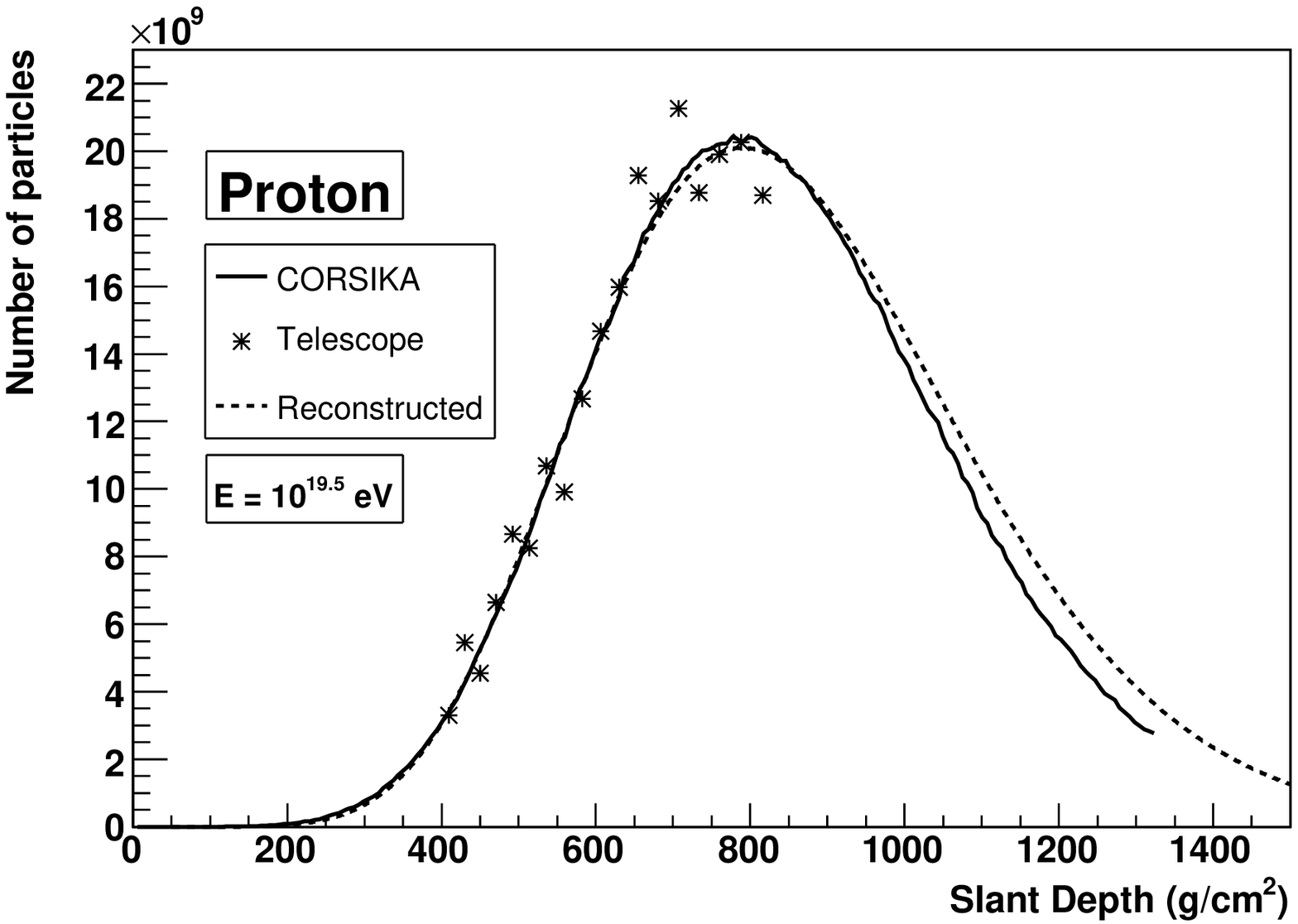}
\end{center}
\caption{Longitudinal development of a proton shower with energy
$10^{19.5}$~eV as given by CORSIKA (full line), as detected by the
HiRes-II telescope according to our simulation where noise has
been added (points) and as reconstructed by fitting a
Gaisser-Hillas function (dotted line).} \label{fig:long:1}
\end{figure}

\begin{figure}[t]
\begin{center}
\includegraphics[width=10cm]{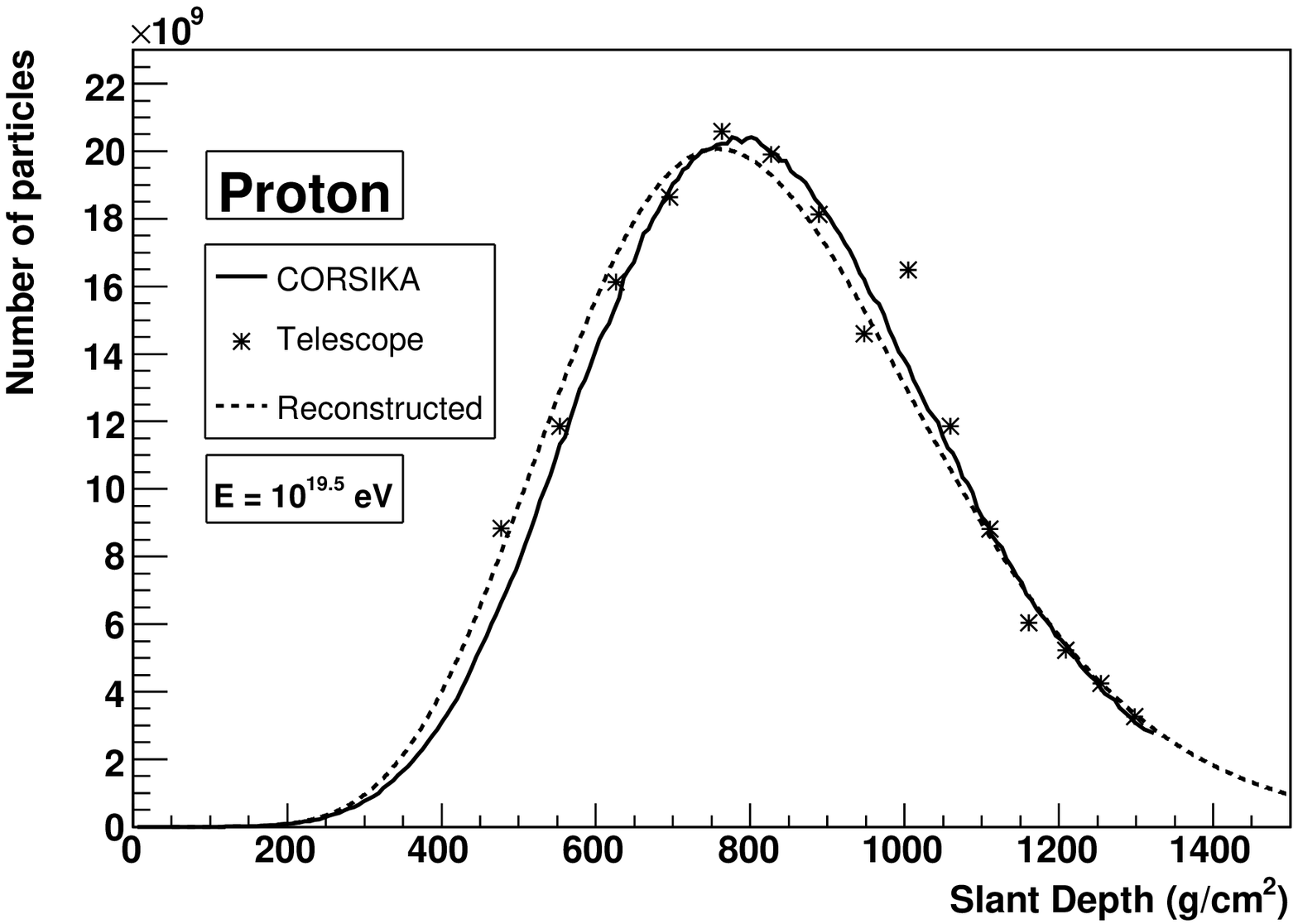}
\end{center}
\caption{Longitudinal development of a proton shower with energy
$10^{19.5}$~eV as given by CORSIKA (full line), as detected by the
HiRes-II telescope according to our simulation where noise has
been added (points) and as reconstructed by fitting a
Gaisser-Hillas function (dotted line).} \label{fig:long:2}
\end{figure}

We try to determine which parameter is less affected by the fact
that the telescopes, in most of the cases, measure only a small
part of the shower, and whether the extrapolation of the observed
longitudinal profile depends on geometric parameters. In order to
do so, we simulated 5,000 CORSIKA proton showers with energy
$10^{19.5}$~eV through the HiRes-II telescopes. For each shower we
compare the integral of the longitudinal number of particles as
calculated by CORSIKA using the whole profile of the shower with
the integral of the fitted longitudinal profile based on the
points actually measured by the telescope. The same comparison is
made between the $N_{\mathrm{max}}$ given by CORSIKA based on the
whole development of the shower and the reconstructed
$N_{\mathrm{max}}$ after the restriction of the shower in the
field of view of the telescope. Fig.~\ref{fig:long:1} and
\ref{fig:long:2} illustrate this procedure. These figures show the
longitudinal profile simulated by CORSIKA in solid black lines.
Star symbol shows the points measured by the telescope after the
inclusion of noise while the dashed line shows the Gaisser-Hillas
fit to the points.

The Gaisser-Hillas fit was implemented according to the
procedures adopted by the HiRes Collaboration in
\cite{bib:hires:espectro} and \cite{bib:hires:espectro:2}.

\begin{figure}[t]
\begin{center}
\includegraphics[width=10cm]{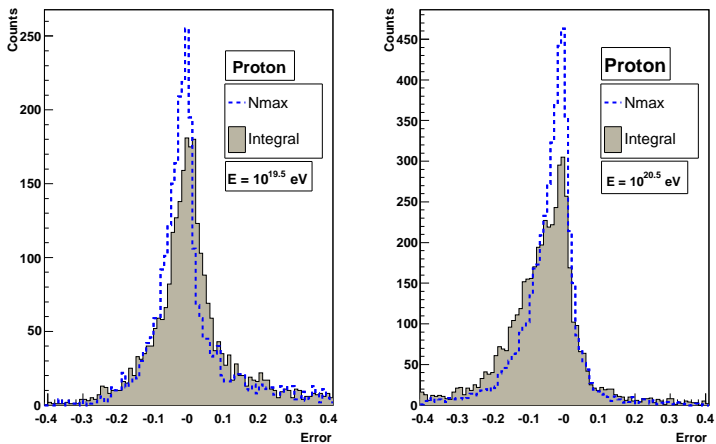}
\end{center}
\caption{Error in the integral and $N_{\mathrm{max}}$
reconstructions due to the fit of a Gaisser-Hillas function using
a limited number of points.} \label{fig:error}
\end{figure}

Fig.~\ref{fig:error} shows the distribution of errors in the
determination of the integral of the longitudinal profile and in
the calculation of the $N_{\mathrm{max}}$ parameter for proton
showers at $10^{19.5}$ and $10^{20.5}$ eV. Table \ref{tab:error}
shows the mean and RMS values for all the distributions. The error
is defined as the reconstructed value minus the simulated value
divided by the simulated value.

%File flutu.cc
% DifInt[i] = (IntRec[i] - IntSim[i])/IntSim[i];
% float DifNmax = (NmaxRec[i] - NmaxSim[i])/NmaxSim[i];

The same behavior is noticed in the $N_{\mathrm{max}}$ and
integral errors. At $10^{19.5}$~eV, the $N_{\mathrm{max}}$
parameter and the integral of the longitudinal profile are
underestimated and the opposite happens at $10^{20.5}$ eV where
both values are overestimated. This might happen because showers
with higher energy develop deeper in the atmosphere hitting ground
level in an earlier stage. However, its is noticeable that the
$N_{\mathrm{max}}$ parameter is less affected by reconstruction
showing a mean closer to zero and a slightly smaller RMS value for
both energies.

The RMS values shown in Table~\ref{tab:error} are the intrinsic
fluctuations of the determination of the $N_{\mathrm{max}}$
parameter and of the integral value of the longitudinal profile.
Nevertheless, we have not applied any quality cuts to the showers
analyzed so far, besides asking that $X_{\mathrm{max}}$ falls
inside the field of view of the telescope.

It has been claimed that $N_{\mathrm{max}}$, as an energy
estimator, should be superior to the integral method for showers
with short track length detected by fluorescence telescopes. Other
authors have also mentioned the dependence of the method on the
shower zenith angle and on the distance from the telescope to the
shower core. However, both zenith angle and distance to the core,
are the variables which determine the track length of the shower
in the field of view of the telescope.

As illustrated in Figs.~\ref{fig:long:1} and \ref{fig:long:2} the
path length is the parameter which influences directly the fit of
the Gaisser-Hillas function and, therefore, the energy
reconstruction. The distance between telescope and shower also
influences the energy reconstruction via the lateral size of the
shower in the photomultiplier camera, as discussed in
\cite{bib:corsika-astro} and \cite{bib:astroph:1dX3d}, but this
effect should be small compared to the reduction of the path seen
by the telescope.

\begin{figure}[t]
\begin{center}
\includegraphics[width=10cm]{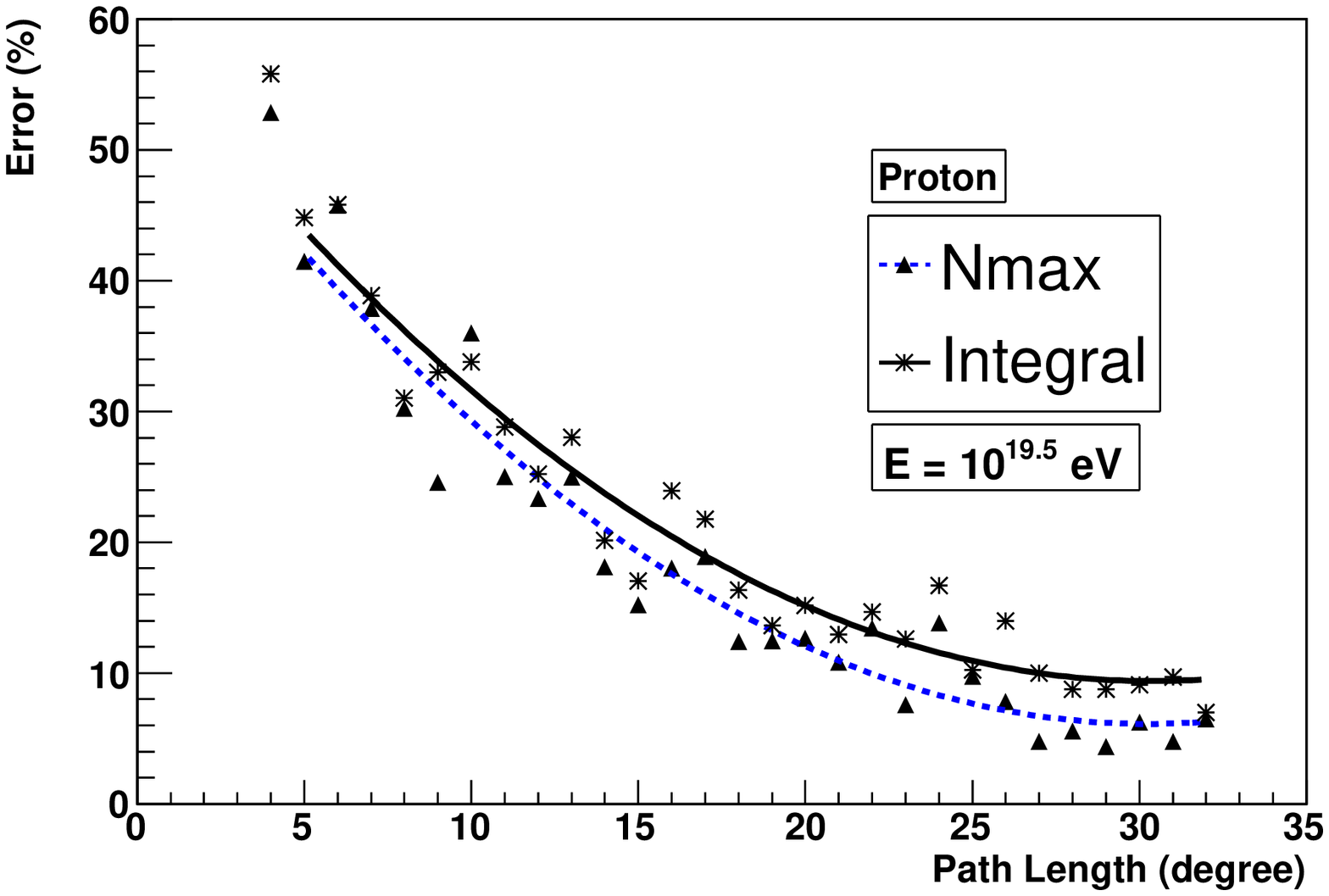}
\end{center}
\caption{Error in the integral and $N_{\mathrm{max}}$ reconstruction due to the fit of a
Gaisser-Hillas function using a limited number of points as a function of the path length
for proton shower with energy $10^{19.5}$~eV. Lines shows polynomial fit to the points in
the range from 5 to 32 degrees.}
\label{fig:erro:path:195}
\end{figure}

Fig.~\ref{fig:erro:path:195} illustrates the reconstruction error
of the integral and of the $N_{\mathrm{max}}$ parameter as a
function of path length for proton showers with energy $10^{19.5}$~eV. 
The line is a polynomial fit to the points. In this case, a
slight better resolution for the $N_{\mathrm{max}}$ parameter is
seen for the entire range of path length.

\begin{figure}[t]
\begin{center}
\includegraphics[width=10cm]{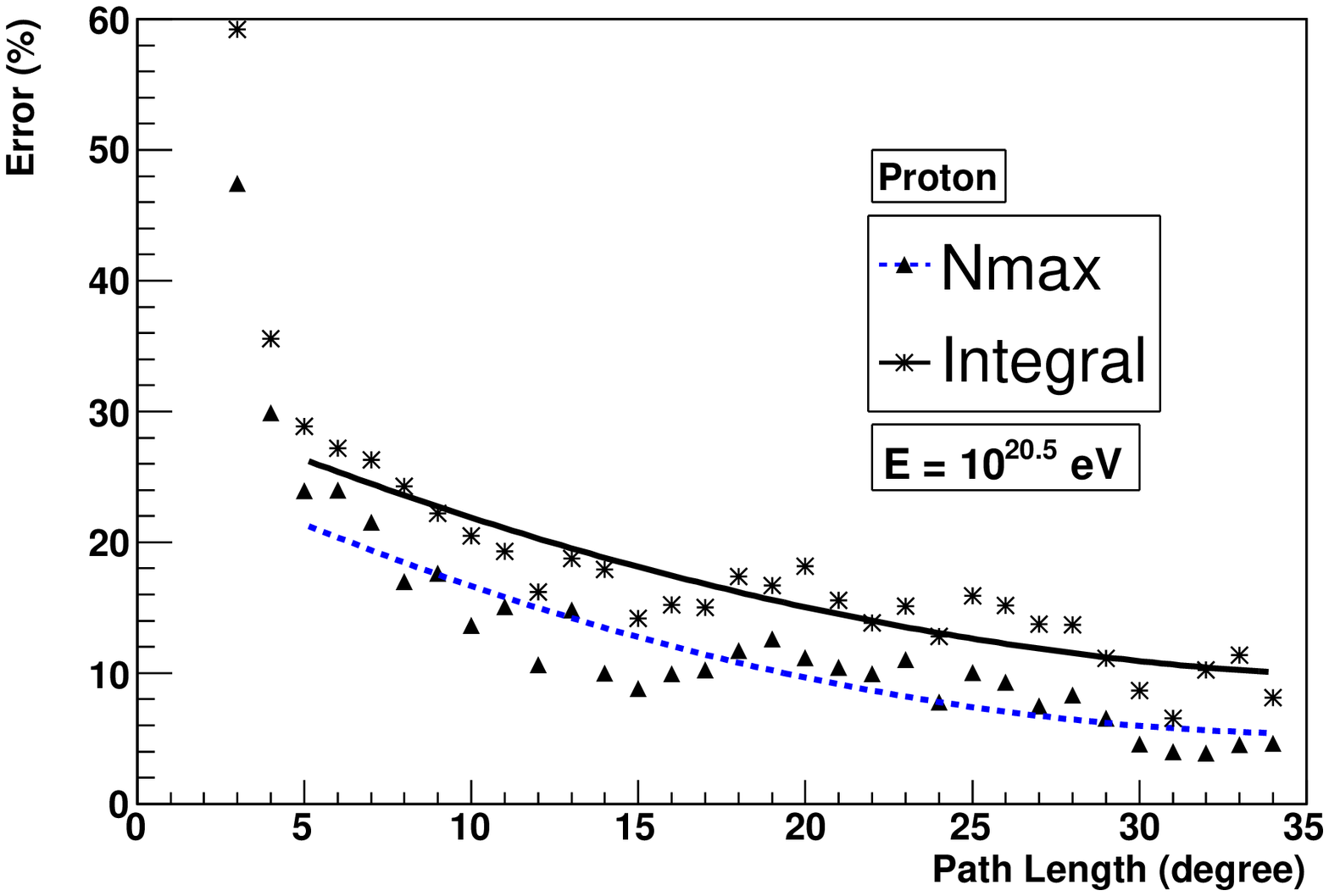}
\end{center}
\caption{Error in the integral and $N_{\mathrm{max}}$ reconstruction due to the fit of a
Gaisser-Hillas function using a limited number of points as a function of the path length
for proton shower with energy $10^{19.5}$~eV.  Lines shows polynomial fit to the points
in the range from 5 to 35 degrees.}
\label{fig:erro:path:205}
\end{figure}

Fig.~\ref{fig:erro:path:205} shows the same plot for proton showers at
$10^{20.5}$ eV where a small difference between the reconstruction of
the $N_{\mathrm{max}}$ and the integral can be seen for the range of
path length from  5 to 35 degrees.

These results show that the reconstruction of the
$N_{\mathrm{max}}$ parameter is equivalent to the reconstruction
of the integral independently of the size of the shower detected
by the telescope. Our calculations suggest that the reconstruction
of the $N_{\mathrm{max}}$ parameter might become slightly more
accurate than the integral reconstruction for the highest energies
above $\sim 10^{20}$ eV.

%##############################################################################
\begin{table}[t]
\begin{center}
\begin{tabular}{|c|c|c|c|c|} \hline
Energy (eV)  &  \multicolumn{2}{|c|}{$N_{\mathrm{max}}$} &  \multicolumn{2}{|c|}{Integral} \\ \hline
            &   Mean  &   RMS   &  Mean  & RMS  \\ \hline
$10^{19.5}$ &   0.02  &   0.16  &  0.03  & 0.17 \\ \hline
$10^{20.5}$ &  -0.04  &   0.10  &  -0.06 & 0.14 \\ \hline
\end{tabular}
\caption{Mean and RMS values for the error distributions given in Fig.~\ref{fig:error}.}
\label{tab:error}
\end{center}
\end{table}
%##############################################################################

%##############################################################################
%##############################################################################
%##############################################################################
%##############################################################################
%##############################################################################
%##############################################################################
%##############################################################################
%##############################################################################
%##############################################################################
%##############################################################################

\section{Relation between primary energy and $N_{\mathrm{max}}$}

Finally, we used the simulation programs already described to
determine the relationship between the reconstructed
$N_{\mathrm{max}}$ and the primary energy of the shower. We used
the same set of CORSIKA showers described above and simulated them
through the HiRes-II and EUSO telescopes. The present are intended
to assess the potential of the $N_{\mathrm{max}}$ parameter as an
energy estimator by means of a full simulation and reconstruction
chain including all types of systematic and statistical errors.

The HiRes-II and EUSO telescopes were chosen to illustrate the
method for ground and space based telescopes. The HiRes-II
telescope is, in principle, very similar to the Auger telescopes
regarding the variables explored here. Both experiments have 30
degrees of field of view in  zenith and the pixel size for the
Auger telescope is 1.5 degree while for the HiRes-II telescope is
1 degree. These small differences should not affect our
conclusions.

\section{HiRes Telescope}

The reconstruction of the shower was also carried out according to
the explanations given by the HiRes Collaboration in
\cite{bib:hires:espectro} and \cite{bib:hires:espectro:2} of their
own procedures. The $X_0$ and $\lambda$ parameters of the
Gaisser-Hillas were fixed to -60 and 70~g/cm$^2$ respectively.

The following quality cuts were applied to the showers after
reconstruction:

%##############################################################################
\begin{itemize}
\item Angular speed $<$ 11
\item Selected tubes $\leq$ 7.
\item 0.85 $<$ Tubes/degree $<$ 3.0.
\item Photoelectrons/degree $>$ 25.
\item Track length $> 7^\circ$, or $>$ $10^\circ$ for events extending above $17^\circ$ elevation.
\item Zenith angle $< 60^\circ$.
\item $150 < X_{\mathrm{max}} < 1200 \ \ \mathrm{g/cm}^{2}$, and is visible in detector
\item Geometry $\chi^{2}/n.d.f. < 10$.
\item Profile fit $\chi^{2}/n.d.f. < 10$.
\end{itemize}
%##############################################################################

\begin{figure}[t]
\begin{center}
\includegraphics[width=10cm]{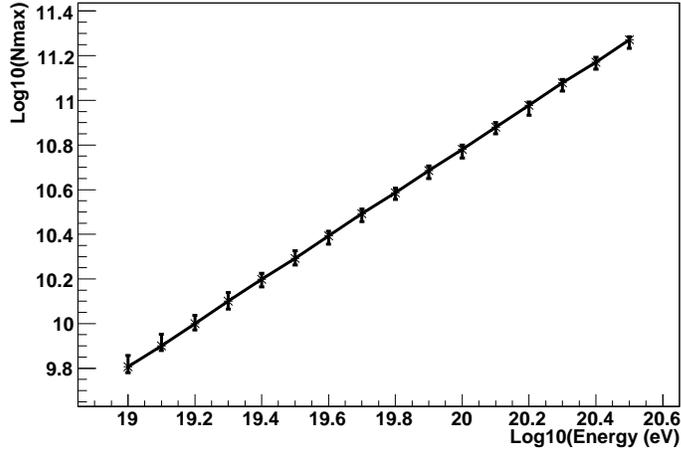}
\end{center}
\caption{Relation between the reconstructed $N_{\mathrm{max}}$ and
the primary energy for showers initiated by protons. Results shown
for the HiRes-II telescope.} \label{fig:en:nmax:hiresII:pr}
\end{figure}

At each energy 5,000 showers were simulated and $N_{\mathrm{max}}$
reconstructed. Fig.~\ref{fig:en:nmax:hiresII:pr} illustrates the
relation between the median $N_{\mathrm{max}}$ and the primary
energy. The error bars represent the 68\% of confidence level of the
reconstructed $N_{\mathrm{max}}$ distribution. The relationship
found is almost linear and can be well fitted by the equation
Log$_{10}$($N_{\mathrm{max}}$) = $-8.75 + 0.97 \times$Log$_{10}$($E$).

\begin{figure}[t]
\begin{center}
\includegraphics[width=10cm]{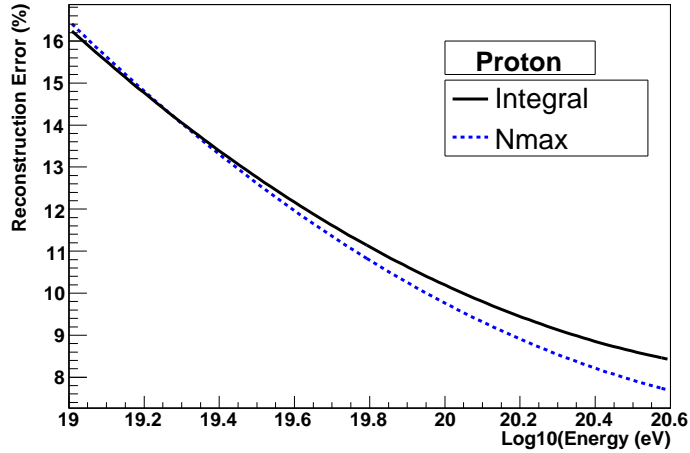}
\end{center}
\caption{Error in the energy reconstruction by the integral and $N_{\mathrm{max}}$
methods as a function of the primary energy. Results shown for the HiRes-II telescope.}
\label{fig:erro:en:pr}
\end{figure}

In order to test the reconstruction of the energy based on the
$N_{\mathrm{max}}$ parameter we simulated a second set of 5,000
independent showers. The energy was reconstructed using the
previous equation and by integrating the Gaisser-Hillas function.
Fig.~\ref{fig:erro:en:pr} presents the error in the energy
reconstruction as a function of the primary energy. As suggested
in the calculations above, the $N_{\mathrm{max}}$ estimator is as
good as the integral procedure for energies between $10^{19}$ and
$10^{19.5}$ eV. For energies above $10^{19.5}$ eV the
$N_{\mathrm{max}}$ estimator gets slightly superior to the
standard integral method.

\begin{figure}[t]
\begin{center}
\includegraphics[width=10cm]{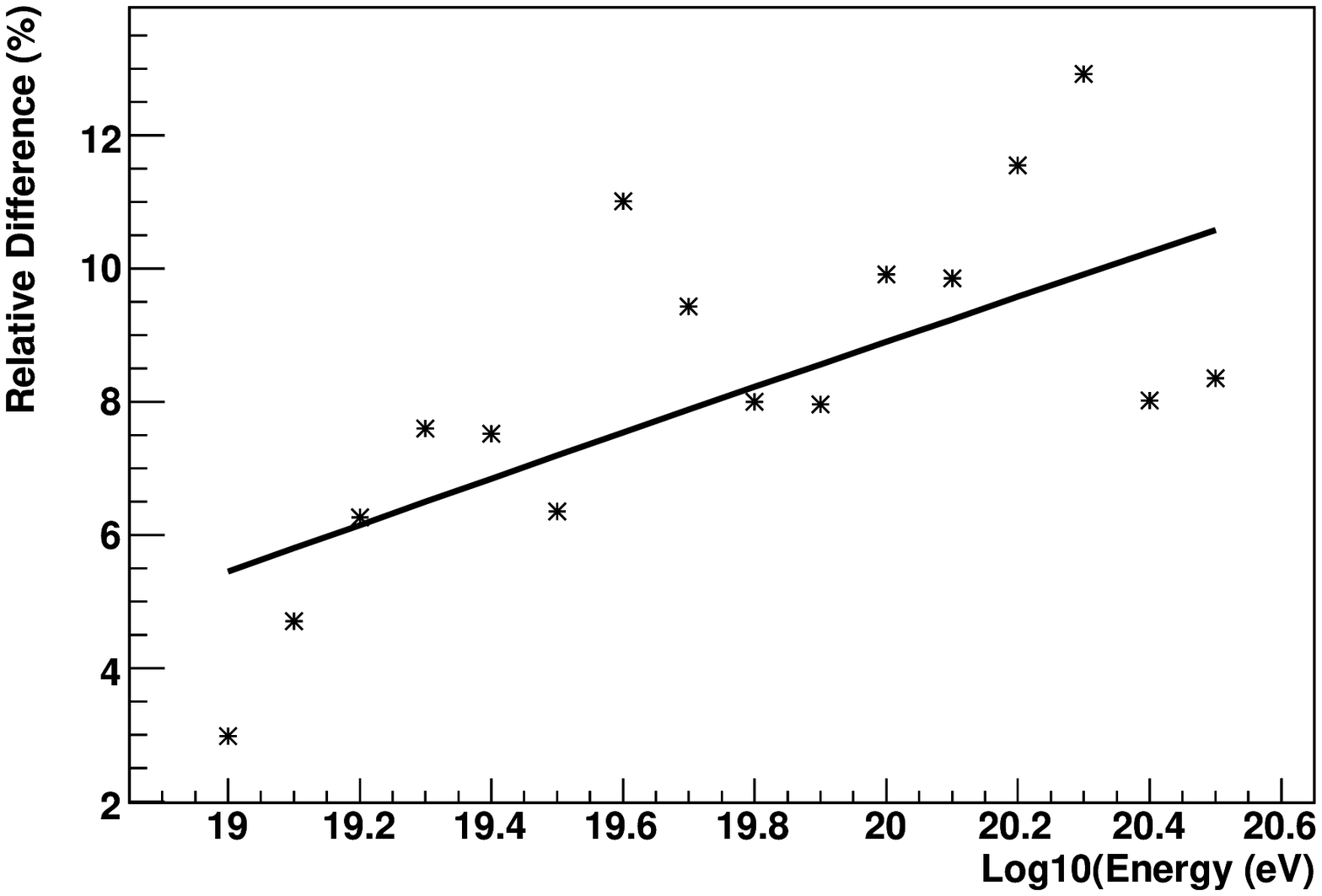}
\end{center}
\caption{Relative difference between the median $N_{\mathrm{max}}$ for proton and gamma
showers. Points shows the difference calculated with the median of the reconstructed
$N_{\mathrm{max}}$ distribution and the solid line shows the difference calculated with
the fitted relation between energy and $N_{\mathrm{max}}$. Note that the line is not a
linear fit to the points.}
\label{fig:rela:diff:pr:ga}
\end{figure}

We also studied the relationship between $N_{\mathrm{max}}$ and
the primary energy for primary gamma showers. The relation is also
linear and very similar to the one for protons shown in
Fig.~\ref{fig:en:nmax:hiresII:pr}. Fig.~\ref{fig:rela:diff:pr:ga}
shows the relative difference between the median
$N_{\mathrm{max}}$ for proton and gamma showers. The difference is
defined as $(N_{\mathrm{max}}^{Pr} -
N_{\mathrm{max}}^{Ga})/N_{\mathrm{max}}^{Pr}$. The points in
Fig.~\ref{fig:rela:diff:pr:ga} show the difference calculated with
the median $N_{\mathrm{max}}$ given by the reconstructed
$N_{\mathrm{max}}$ distribution at each energy for proton and
gamma primaries. The solid line is the difference calculated using
the fitted relations between energy and $N_{\mathrm{max}}$.

For the energy range $10^{19.0}$ to $10^{20.5}$ eV, the difference
is smaller than 12\%. This difference would influence the energy
reconstruction in a mixed composition of proton and gamma
primaries. The same effect would be seen in both reconstruction
methods: $N_{\mathrm{max}}$ and integral according to the
discussion in Section \ref{sec:fluc}.

If a mixed composition of 50\% proton and 50\% gammas is
considered the energy resolution would degrade but still would
show a behavior similar to the one show in
Fig.~\ref{fig:erro:en:pr}. At $10^{19}$ eV the resolution would
degrade to 20\% for both methods. At $10^{20.5}$ the resolution
for the $N_{\mathrm{max}}$ method would be 16\% and for the
integral procedure it would be around 18\%.

%##############################################################################
%##############################################################################

%##############################################################################
%##############################################################################

%##############################################################################
%##############################################################################

\section{EUSO}

The same analysis were extended to the EUSO telescope. The details
of this experiment were included in the simulations as explained
in Section \ref{sec:sim}. However, not only the telescope is
different but its peculiar field of view also requires a
considerable change in the shower reconstruction program. This can
be verified in Fig.~\ref{fig:euso:accep} in which we show the
EUSO acceptance.

Since there are no quality cuts yet optimized for the EUSO
telescope, we imposed very loose ones requiring total path length
greater than $0.6^\circ$ and greater than 200~g/cm$^2$ and
$X_{\mathrm {max}}$ in the field of view of the telescope. We
would like to stress that different quality cuts will modify the
relation between $N_{\mathrm{max}}$ and energy presented here.

\begin{figure}[t]
\begin{center}
\includegraphics[width=10cm]{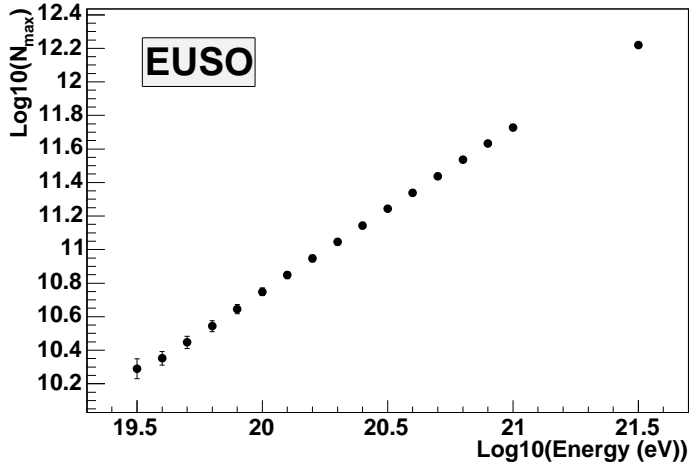}
\end{center}
\caption{Relation between the reconstructed $N_{\mathrm{max}}$ and
the simulated primary energy for showers initiated by protons.
Results are for the EUSO telescope.} \label{fig:nmax:en:euso}
\end{figure}

Fig.~\ref{fig:nmax:en:euso} shows the relation between the
reconstructed $N_{\mathrm{max}}$ and the simulated energy, which
can be well approximated by the equation: 
Log$_{10}$($N_\mathrm{max}$) = -8.95 + 0.90 $\times$ Log$_{10}(E)$.

\begin{figure}[t]
\begin{center}
\includegraphics[width=10cm]{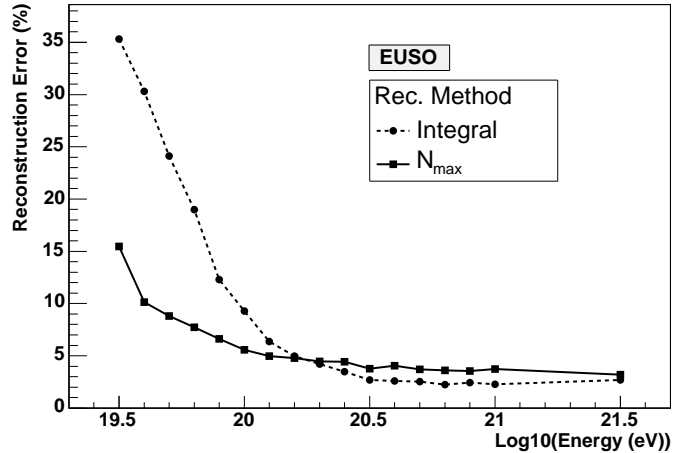}
\end{center}
\caption{Error in the energy reconstruction by the integral and
$N_{\mathrm{max}}$ methods as a function of the primary energy.
Results are for the EUSO telescope.} \label{fig:erro:en:euso}
\end{figure}

In order to test the reconstruction of the energy based on
$N_{\mathrm{max}}$ we simulated a second set of 5,000 independent
shower. The energy was reconstructed using both, the equation
above and the integration of the Gaisser-Hillas fitted profile.
Fig.~\ref{fig:erro:en:euso} shows the error in the energy
reconstruction as a function of the primary energy. For energies
below $10^{20.2}$~eV, the $N_{\mathrm{max}}$ estimator was superior
to the integral method. For energies above $10^{20.2}$~eV, both
methods are equivalent.

These results are different from those for the HiRes-II
telescopes, which confirms that the relation between
$N_{\mathrm{max}}$ and primary energy depends on the telescopes
and details of the analysis, mainly the quality cuts.

The exact influence of each quality cut on the energy resolution
is not completely understood at present and is out of the scope of
this paper. However, the results shown here suggest that the
quality cuts imposed by the HiRes Collaboration are well optimized
in order to guarantee a good energy resolution for the integral
method for the energy ranging from $10^{19}$ to $10^{20.5}$ eV. On
the other hand, the loose cuts applied by us to the EUSO analysis
do not achieve an optimal selection of event for energies below
$10^{20.2}$ eV resulting in an increase of the reconstruction
error for both techniques. It is also apparent that
$N_{\mathrm{max}}$, for EUSO, is less sensitive to well calibrated
cuts because, despite the fact that the errors in this method
increases at lower energies, they still remain within reasonable
bounds, i.e., smaller than 15\% down to  $10^{19}$ eV.

%##############################################################################
%##############################################################################

%##############################################################################
%##############################################################################

\section{Conclusion}

In the present work, we tested the number of particles at the maximum air 
shower development, $N_{\mathrm{max}}$, as an energy estimator. The
uncertainties associated with the simulation models were calculated and 
compared to the uncertainties in the missing energy calculation. A similar 
intrinsic fluctuation of around 5-10\% at $10^{19.5}$~eV was verified for 
both approaches. These uncertainties decrease with increasing energy.

The systematic uncertainties in the determination of $N_{\mathrm{max}}$ 
are dominated by mass composition rather than hadronic interaction models 
if gamma showers are taken into account. If gamma showers are excluded 
from the analysis, mass composition and hadronic interaction models would 
contribute equivalently to the total uncertainties.

Full telescope simulations were considered in our analysis. The
first source of uncertainty in the reconstruction procedure is the
fit of the points measured by the telescope using a Gaisser-Hillas
function. We investigated the statistical uncertainties involved
in the evaluation of $N_{\mathrm{max}}$ and of the integral of the
longitudinal distribution function as a function of the path
length seen by the telescope. $N_{\mathrm{max}}$ appears to be
slightly better reconstructed independent from the path length
observed by the telescope. At $10^{19.5}$ eV the difference in
reconstructing $N_{\mathrm{max}}$ and the integral is very small,
however with increasing energy, at $10^{20.5}$ eV for example, the
difference is of the order of a few percent, depending on energy
(see Fig.~\ref{fig:erro:path:205}).

The relation between the $N_{\mathrm{max}}$ and primary energy for
proton showers was calculated for the HiRes-II telescope including
the quality cuts used by the HiRes collaboration.
$N_{\mathrm{max}}$ appears to be equivalent to the standard
integral procedure as an energy estimator, with a small
superiority of the $N_{\mathrm{max}}$ approach for the highest
energies (see Fig.~\ref{fig:erro:en:pr}). We also calculated the
same relation for gamma showers and the relative difference for
the proton shower is smaller than 12\%. If a mixed composition of
50\% proton and 50\% gammas is considered the energy resolution
degrades to 20\% at $10^{19}$ eV for both methods. At $10^{20.5}$
the resolution for the $N_{\mathrm{max}}$ method is 16\% while for
the integral procedure is $\sim 18\%$.

The same analysis were performed for the EUSO telescope and we
demonstrated that the $N_{\mathrm{max}}$ method also works well
for space based telescopes. The specific relationships, however,
depend on the telescopes characteristics and reconstruction
methods.

Finally, $N_{\mathrm{max}}$ was established as a good energy
estimator for fluorescence telescope with some advantages over the
standard integral procedure.

\section{Acknowledgments}

This work was supported by the Brazilian science foundations
FAPESP  and CNPq. F. Sanches thanks IAG/USP for the hospitality
and INFN/UNIMI for funding supports. Most simulations were carried out
on a Cluster 
Linux TDI, supported by Laborat\'orio de Computa\c c\~ao
Cient\'{\i}fica Avan\c cada at Universidade de S\~ao Paulo.

\bibliographystyle{elsart-num}
\bibliography{nmax_ref.bib}

\end{document}